\documentclass[twocolumn, superscriptaddress, nofootinbib, preprintnumbers, prl]{revtex4}

\usepackage{CJK}
\usepackage{tikz}
\usepackage{amssymb}
\usepackage{epsfig}
\usepackage{bbm}
\usepackage{bbold}
\usepackage{graphicx}
\usepackage{hyperref}
\usepackage{romannum}
\usepackage{graphics,graphpap}
\usepackage{graphicx}
\usepackage{amsmath,exscale,amsthm}
\usepackage{graphicx}
\usepackage{yfonts}
\usepackage{mathrsfs}
\usetikzlibrary{trees}
\usetikzlibrary{positioning,arrows}
\usetikzlibrary{decorations.pathmorphing}
\usetikzlibrary{decorations.markings}
\usetikzlibrary{decorations.text}
\usetikzlibrary{fit,chains,calc,shapes.geometric,intersections}

\begin{document}

\def\a{\alpha}
\def\b{\beta}
\def\c{\chi}
\def\d{\delta}
\def\e{\epsilon}
\def\f{\phi}
\def\g{\gamma}
\def\h{\eta}
\def\i{\iota}
\def\j{\psi}
\def\k{\kappa}
\def\la{\lambda}
\def\m{\mu}
\def\n{\nu}
\def\o{\omega}
\def\p{\pi}
\def\q{\theta}
\def\r{\rho}
\def\s{\sigma}
\def\t{\tau}
\def\u{\upsilon}
\def\x{\xi}
\def\z{\zeta}
\def\D{\Delta}
\def\F{\Phi}
\def\G{\Gamma}
\def\J{\Psi}
\def\L{\Lambda}
\def\O{\Omega}
\def\P{\Pi}
\def\Q{\Theta}
\def\S{\Sigma}
\def\U{\Upsilon}
\def\X{\Xi}

\def\ve{\varepsilon}
\def\vf{\varphi}
\def\vr{\varrho}
\def\vs{\varsigma}
\def\vq{\vartheta}

\def\dg{\dagger}                                     % hermitian conjugate
\def\ddg{\ddagger}                                   % double dagger
\def\wt#1{\widetilde{#1}}                    % big tilde
\def\mt{\widetilde{m}_1}
\def\mti{\widetilde{m}_i}
\def\mtj{\widetilde{m}_j}
\def\rt{\widetilde{r}_1}
\def\mtt{\widetilde{m}_2}
\def\mttt{\widetilde{m}_3}
\def\rtt{\widetilde{r}_2}
\def\mb{\overline{m}}
\def\VEV#1{\left\langle #1\right\rangle}        % < >
\def\be{\begin{equation}}
\def\ee{\end{equation}}
\def\ds{\displaystyle}
\def\ra{\rightarrow}

\def\bea{\begin{eqnarray}}
\def\eea{\end{eqnarray}}
\def\NO{\nonumber}
\def\Bar#1{\overline{#1}}
\def\ylz{\textcolor{red}}

\begin{CJK*}{GB}{} 

\pagenumbering{arabic}
%Title of paper
\title{Gravitational waves from neutrino mass and dark matter genesis}

\author{Pasquale Di Bari}
\affiliation{{\it\small School of Physics and Astronomy},
{\it\small University of Southampton,} 
{\it\small  Southampton, SO17 1BJ, UK} }
\author{Danny Marfatia}
\affiliation{{\it\small Department of Physics and Astronomy},
{\it\small University of Hawaii at Manoa,} 
{\it\small  Honolulu, HI 96822, USA} } 
\author{Ye-Ling Zhou}

\affiliation{{\it\small School of Physics and Astronomy},
{\it\small University of Southampton,} 
{\it\small  Southampton, SO17 1BJ, UK} }

\begin{abstract}
We introduce a model in which the genesis of dark matter (DM) and neutrino masses is associated with a first order 
phase transition of a scalar singlet field.  During the phase transition a source right-handed neutrino (RHN)
acquires a spacetime-dependent mass dynamically, a small fraction of which is converted via resonant oscillations
into a very weakly mixed dark RHN which decays to a dark matter RHN with the observed relic abundance. 
Neutrino masses are generated via a traditional two RHN type-I seesaw 
between a fourth RHN and the source neutrino. 
The gravitational waves produced during the phase transition have a peak frequency 
that increases with the DM mass, 
and are detectable at future gravitational wave interferometers
for DM masses above $\sim 1\,{\rm MeV}$. 
Since the source RHNs are heavier than the electroweak scale, successful leptogenesis is also attainable.  
\end{abstract}

\maketitle
\end{CJK*}

{\bf 1.~Introduction}. The nature of dark matter (DM) is one of the longest-standing puzzles in 
fundamental physics~\cite{bertonehooper}. Although astrophysical and cosmological observations
support a solution in terms of a new particle and disfavor alternative explanations in terms
of modified gravity or primordial black holes, all efforts to identify the nature of this new particle,
with direct, indirect and collider searches, have failed so far. This produces strong constraints
on the existing models of DM, either favoring heavy DM particles (with mass above a TeV), or light particles (with mass below a GeV) 
or very weakly coupled ones, such as axion-like particles, or some combination thereof. 
Therefore, a non-thermal production mechanism, 
quite different from the usual WIMP paradigm, that relies on 
very small couplings of the DM particle to the thermal bath,
is a reasonable possibility to consider.

At the same time extensions of the standard model (SM) should also account for neutrino masses and mixing, and explain the matter-antimatter asymmetry of the universe.  The type-I seesaw mechanism is the most minimal and attractive way to incorporate neutrino masses and mixing and to explain the matter-antimatter asymmetry of the universe via leptogenesis. It is then quite reasonable to seek unified models of neutrino masses, DM and leptogenesis starting from the type-I seesaw Lagrangian and extended with some new ingredient that also addresses DM. 

An example of this kind was proposed in Ref.~\cite{ad} in which one  (dark) right-handed neutrino (RHN) has vanishing Yukawa couplings and therefore does not contribute to neutrino masses and mixing.  However, it mixes with the other (source) RHNs and to the standard model Higgs through the 5-dimensional non renormalizable operator 
$({\lambda_{IJ} / \Lambda})\Phi^\dagger \, \Phi \, \overline{N^c_{R I}}\, N_{R J}$~\cite{anisimov}.
Consequences of these interactions have been studied in Refs.~\cite{unified,densitym}.

In this Letter we introduce a novel mechanism for the production of RHN DM that relies on a first order 
phase transition of a scalar singlet field.
It has been proposed that the variation of couplings in the Weinberg operator during a phase transition can provide a new way to implement leptogenesis~\cite{yelingjessicasilvia}.  
Here we show that a phase transition can induce an efficient
conversion of source RHNs into a dark RHN species that decays prior to the onset of big bang nucleosynthesis, into a lighter dark RHN
species that plays the role of DM.  The process is compatible with a standard cosmological history. 
A strong first order phase transition is required and an associated production of gravitational
waves (GWs) is expected~\cite{witten,hogan,tw,kkt}. In fact the spectrum of GWs is related to the 
properties of the DM particle in a way that provides an interesting signature of the model. 

Several works have proposed a more or less direct link of DM genesis to a phase transition~\cite{witten,Cohen:2008nb, Falkowski:2012fb,Huang:2017kzu,Bian:2018mkl,Bai:2018dxf,murayama,kopp,Chway:2019kft,tseng},
some of which also predict a detectable GW spectrum. Additionally, in our scenario
there is also an important link with neutrino masses and leptogenesis.

{\bf 2.~The model}. In addition to the SM particle content we have four RHNs $N_{\rm DM}$, $N_{\rm D}$, $N_1$, $N_2$
and a new complex scalar field $\eta$ which may be associated with the breaking of a flavor symmetry.  
The two dark RHNs, $N_{\rm DM}$ and $N_{\rm D}$, are  neutral gauge singlets which acquire a Majorana mass 
$M_{\rm DM}$ and $M_{\rm D}$ respectively, with $M_{\rm DM} < M_{\rm D}$,  by e.g., symmetry breaking within
a dark sector; $N_{\rm DM}$ is the DM candidate. In a similar way, a source RHN $N_{\rm S}$, either $N_1$ or $N_2$, 
acquires a mass by coupling to a different scalar field. 
The important thing is that the dark sector is weakly coupled to the visible sector
 through a small mixing of the dark RHN $N_{\rm D}$ with $N_{\rm S}$ 
playing the role of a source RHN~\cite{ad}.  
This mixing is generated through Higgs portal interactions of $N_{\rm D}$ with $N_{\rm S}$
and the SM Higgs $\Phi$, and is described by the operator,
$(\la^{\rm mix}_{\rm DS}/\Lambda_{\rm DS})\,\Phi^\dagger \, \Phi \, \overline{N_{\rm D}^c} \,{N_{\rm S}}$~\cite{anisimov,ad,unified}. We also 
introduce a similar operator describing the mixing between the two dark RHNs,
$(\la^{\rm mix}_{\rm DD}/\Lambda_{\rm DD})\,\Phi^\dagger \, \Phi \, \overline{N_{\rm D}^c} \,{N_{\rm DM}}$, which is expected to be generated
at a different energy scale $\L_{\rm DD}$ and with a different coupling $\la^{\rm mix}_{\rm DD}$. For simplicity, we assume that the same new physics generates 
both operators and that the mixing is the same: $\Lambda_{\rm DS}/\la^{\rm mix}_{\rm DS} = \L_{\rm DD}/\la^{\rm mix}_{\rm DD} = \widetilde{\L}$.
The source RHN $N_{\rm S}$ also has  a coupling $\la_{\rm S}$ to the new scalar field $\eta$. 
Here we do not specify the ultraviolet complete model that produces such a mixing
but the most attractive option is that the same physics is responsible for both
$\la^{\rm mix}_{\rm D(D,S)}$ and $\la_{\rm S}$.
Thus, the SM is extended by
 \bea\label{mixingPT}
-{\cal L}_{\la} & = & {1\over 2}\,M_{\rm DM}\, \overline{N_{\rm DM}^c} \, N_{\rm DM} + {1\over 2}\,M_{\rm D}\, \overline{N_{\rm D}^c} \, N_{\rm D} + {\lambda_{\rm S}\over 2}  \, \eta \, \overline{N_{\rm S}^c} \, N_{\rm S} \nonumber\\ 
&  & + {1\over \widetilde{\L}}\,\Phi^\dagger \, \Phi \, \overline{N_{\rm D}^c} \, N_{\rm S}
+{1\over \widetilde{\L}}\,\Phi^\dagger \, \Phi \, \overline{N_{\rm DM}^c} \, N_{\rm D}
+ {\rm h.c.} \,  .
\eea
The scalar field  $\eta$ acquires a vev $v_{\eta}$ during a first order phase
transition and, simultaneously,
the source RHN acquires a spacetime-dependent mass. 

Including a fourth RHN, one also recovers 
the usual type-I seesaw Lagrangian with two RHNs
that describes neutrino masses and mixing. At the end of the
phase transition the Lagrangian terms extending the SM are (with $I,J = 1,2$)
\bea\label{finalagrangian}
-{\cal L}& = & \overline{L_{\a}}\,h_{\a J}\, N_{J}\, \widetilde{\Phi} \nonumber
                          + \frac{1}{2} \, M_{I} \, \overline{N^{c}_{ I}} \, \delta_{IJ} \, N_{ J}    \\ 
                       & + &   {1\over \widetilde{\L}}\,\Phi^\dagger \, \Phi \, \overline{N_{\rm DM}^c} \, N_{\rm D}  
                        + {1\over \widetilde{\L}}\,\Phi^\dagger \, \Phi \, \overline{N_{\rm D}^c} \, N_{\rm S} 
                       \\ \nonumber
                  & + &       {1\over 2}\,M_{\rm DM}\, \overline{N_{\rm DM}^c} \, N_{\rm DM}  + 
                                {1\over 2}\,M_{\rm D}\, \overline{N_{\rm D}^c} \, N_{\rm D}       + \mbox{\rm h.c.}  \,  ,
\eea
where either $N_1$ or $N_2$ is identified with the  source RHN and the other
neutrino is assumed to have negligible or no mixing with the dark neutrinos. Then, the second seesaw RHN
plays no role in DM genesis but, in combination with the source RHN,
determines neutrino masses and mixing via a two RHN seesaw mechanism which can also potentially
generate the matter-antimatter asymmetry via leptogenesis~\cite{fy}.

{\bf 3.~Dark matter genesis}. The production of the dark RHN abundance  can be calculated by
solving the density matrix equation ($I,J=N_D, N_S$)~\cite{densitym}
\bea\label{densitymatrixeq}
{dN_{IJ} \over dt} & = & -i\,[\Delta{\cal H}, N]_{I J}  - 
\begin{pmatrix}
0   &  \Gamma_{\rm dec}  \\ 
\Gamma_{\rm dec} &   \Gamma_{\rm prod}
\end{pmatrix}
\,   ,
\eea
where $N_{IJ}$ is the abundance density matrix containing on the diagonal terms the dark and source
RHN abundances, normalised in a way that they are simply unity in ultra-relativistic thermal equilibrium.
The quantities $\Gamma_{\rm dec}$ and $\Gamma_{\rm prod}$ are  
the decoherence and production rates respectively.  The effective hamiltonian is given by
\be\label{dmeq}
\Delta {\cal H}_{IJ} \simeq   
\left( \begin{array}{cc}
- \frac{\D \widetilde{M}^2}{4 \, p} &  \D H_{\rm mix}   \\[1ex]
\D H_{\rm mix} &  \frac{\D \widetilde{M}^2}{4 \, p} 
\end{array}\right)  \, ,
\ee
where $\D H_{\rm mix} \equiv T^2/(12\,\widetilde{\L})$,
$\D \widetilde{M}^2 \equiv \widetilde{M}^2_{\rm S}(r,t) - M^2_{\rm D}$, with
the $N_{\rm S}$ effective thermal mass given by the sum of 
a spacetime dependent term and thermal terms:
\be\label{M2tildeS}
\widetilde{M}^2_{\rm S}(r,t) = M^2_{\rm S}(r,t) +\frac{T^2}{4} \, h^2_{\rm S} + {T^2 \over 8}\,\la_{\rm S}^2\, N_{N_{\rm S}}\,N_{\eta} \,   .
\ee
In this expression $h^2_{\rm S} \equiv (h^\dagger \, h)_{\rm SS}$ and we introduce the
usual effective neutrino mass, $\widetilde{m}_{\rm S} = v^2\,h^2_{\rm S}/M_{\rm S}$.
The seesaw mechanism requires $\widetilde{m}_{\rm S}/m_{\rm sol} \geq 1$, 
where $m_{\rm sol}\simeq 8.6\,{\rm meV}$ is the solar neutrino mass scale.  
We take the minimum value $\widetilde{m}_{\rm S}/m_{\rm sol} = 1$  that
maximizes the DM lifetime, so that $h^2_{\rm S} = m_{\rm sol}\,M_{\rm S}/v^2$.

The spacetime-dependent mass $M_{\rm S}(r,t) = \la_{\rm S}\,v_{\eta}(r,t)$
is generated by the vev of $\eta$ that is described by a well known
kink solution for the bubble wall profile~\cite{john},
\be
v_{\eta}(r,t) = {1\over 2}\,\bar{v}_{\eta} \,
\left[1-\tanh \left( {r-v_{\rm w}\,(t-t_\star) \over \D_{\rm w} }\right) \right] \,  ,
\ee
where $v_{\rm w}$ and $\D_{\rm w}$ are the bubble wall velocity  and width.
Note that points in the false vacuum with $v_{\eta}=0$ (outside the bubble),
correspond to $r \gg \Delta_{\rm w} + v_{\rm w}(t-t_{\star})$.

We assume that before the phase transition  the  $\eta$ abundance $N_{\eta}$ gets 
thermalized, so that we can set $N_\eta = N_{\eta}^{\rm eq} = N_{\g} = 4/3$ in Eq.~(\ref{M2tildeS}). 
On the other hand, the source RHN abundance $N_{N_{\rm S}}$ is obtained by solving 
Eq.~(\ref{densitymatrixeq}). We adopt a monochromatic approximation with
 $p \simeq 3\,T$. The production rate $\Gamma_{\rm prod}$ for the $N_{\rm S}$ abundance is 
given by the sum of two contributions: $\G_{\rm D}+\G_{\rm S}$, from inverse decays and scatterings from the Yukawa couplings 
to the Higgs, and  $2 \G_{\eta\ra N_{\rm S}N_{\rm S}}$, from the decays of  $\eta$ into source RHNs. Therefore, explicitly we can write
\be\label{rateeqbis}
{dN_{N_{\rm S}} \over dz} \simeq - (D+S) (N_{N_{\rm S}}-N_{N_{\rm S}}^{\rm eq}) 
- D_{\eta}\,N^{\rm eq}_{\eta}\,(N^2_{N_{\rm S}}-(N_{N_{\rm S}}^{{\rm eq}})^2)  ,
\ee
where we neglected a tiny oscillatory term from the Liouville-von Neumann term 
from the mixing with $N_{\rm DM}$. 
We also introduced the variable $z=T_{\star}/T$, where $T_{\star}$ is a convenient 
energy scale that we identify with the temperature of the phase transition, and 
$(D,S,D_\eta) \equiv (\G_{\rm D},\G_{\rm S},2\G_{\eta\ra N_{\rm S}N_{\rm S}})/(H(z)\,z)$ with the expansion rate
\be
H(z) = \sqrt{8\pi^3\,g_{\star} \over 90} \, {T_{\star}^2 \over M_{\rm Pl}} \, {1\over z^2}
\simeq 1.66 \, \sqrt{g_{\star}}\,{T_{\star}^2\over M_{\rm Pl}}\,{1 \over z^2} \,.
\ee
For $T\gtrsim M_{\rm S}$ one has $D+S \simeq 2\, (\widetilde{m}_{\rm S}/m_{\rm sol})\,(M_{\rm S}/T_{\star})\,
(1 + (8\pi^2/9)\,(M_{\rm S}/T_{\star})\,z)$. The term from $\eta$ interactions can be written in the form
$D_{\eta} = (\widetilde{m}_{\rm S}^\eta/(2\, m_\star))\,\left({m_\eta / T}\right)^2$, where 
$m_{\star}\equiv 16\,\pi^{5/2}\, \sqrt{g_{\star}}/(3 \sqrt{5})\,(v^2 / M_{\rm Pl}) \simeq 1.08 \,{\rm meV}$
is the usual equilibrium neutrino mass and, in a similar fashion to Yukawa couplings, we have also introduced 
the effective neutrino mass  $\widetilde{m}_{\rm S}^\eta \equiv v^2\,\lambda^2_{\rm S}/T_{\star}$ that
parameterizes the coupling of the source RHN to the scalar $\eta$.
The scalar $\eta$ has a vanishing or negligible bare mass but it gets a thermal mass $m_{\eta} = g_{\eta}\,T$, where $g_{\eta}$
is the $\eta$ coupling constant to the thermal bath and typical values are $g_{\eta} \sim 0.1$.
For example, for the SM Higgs boson, $m_H/T \simeq 0.4$. 
Henceforth, we set $g_{\eta} = 1/6$.
Since in our case it will turn out that $M_{\rm S} \simeq T_{\star}$, 
Yukawa coupling and $\eta$ interactions thermalize the source neutrinos and $N_{N_{\rm S}} \simeq N_{N_{\rm S}}^{\rm eq} \simeq 1$. 

We now calculate the dark RHN abundance by solving Eq.~(\ref{densitymatrixeq}). 
This can be done by noticing that before the phase transition, $M_{\rm S}(r,t) =0$ in Eq.~(\ref{M2tildeS}).
In this way thermal medium effects dominate and suppress the mixing since the off-diagonal term
$\D H_{\rm mix}$ is negligible compared to the contribution from $\eta$ interactions 
in Eq.~(\ref{M2tildeS}).  During the phase transition, the source RHN mass $M_{\rm S}(r,t)$ increases from zero to its final value $M_{\rm S} = \lambda_{\rm S}\, \bar{v}_{\eta}$. 
Therefore, during the phase transition, there can be a time when 
$\D \widetilde{M}^2(r,t_{\rm res}) = 0$, or equivalently
$\widetilde{M}^2_{\rm S}(r,t_{\rm res}) = M^2_{\rm D}$, 
corresponding to a mixing resonance condition
between $N_{\rm S}$ and $N_{\rm D}$:
\bea \label{MDMMS} 
{M^2_{\rm D}\over M^2_{\rm S}} \simeq 
\left[\frac{1}{2} - \frac{1}{2}\tanh \left( {r-v_{\rm w}\,(t_{\rm res}-t_{\star}) \over 
\D_{\rm w} }\right)\right]^2+{T_{\star}^2 \over 6 \bar{v}_{\eta}^2},
 \eea
where   
we neglected the effective potential generated by the standard Yukawa coupling term 
since it is much smaller than the term generated by the $\eta$ interactions.  
Typically, $T_{\star} \sim \bar{v}_{\eta}$ and, for definiteness, we fixed $T_{\star}/\bar{v}_{\eta}=1$.
The resonance condition (\ref{MDMMS}) shows that, for given values of $T_{\star}$ and $M_{\rm S}$, 
there is a finite range,  $M_{\rm D}^{\rm min} \leq M_{\rm D} \leq M_{\rm D}^{\rm max}$, 
corresponding to a variation of the hyperbolic tangent within $[-1,1]$. For given
 values of $M_{\rm D}$ and $M_{\rm S}$, the resonance can occur either inside or outside the bubble wall, 
depending on whether $|\tanh\left[( {r-v_{\rm w}\,(t_{\rm res}-t_{\star}))/ \D_{\rm w} }\right]| \ll 1$ or $\simeq 1$.
Therefore, for a fixed value of $T_{\star}$,
the resonance condition  (\ref{MDMMS}) identifies an allowed region in the ($M_{\rm S}$, $M_{\rm D}$) plane.
\begin{figure}
        \psfig{file=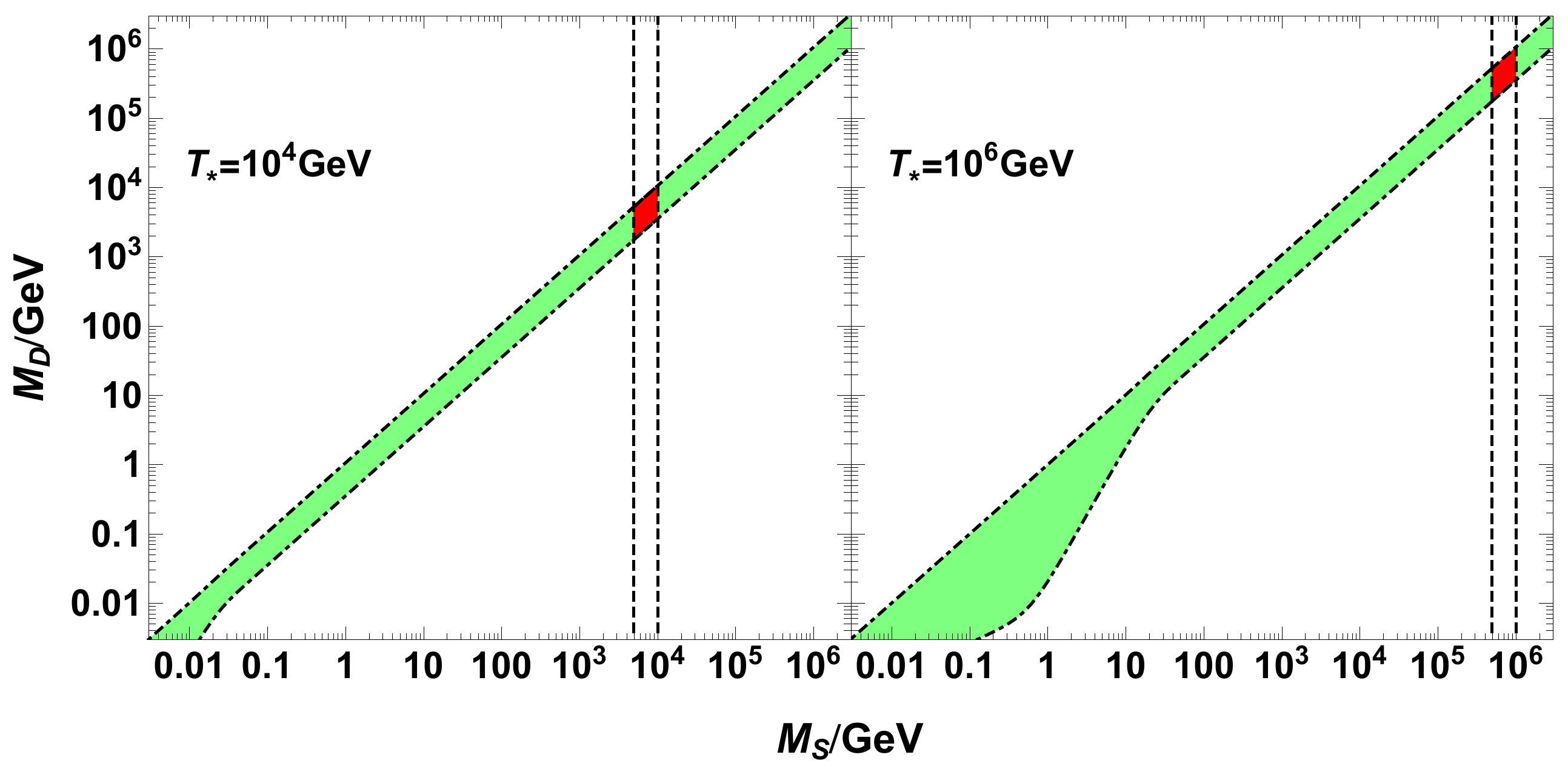,height=45mm,width=80mm} \\
\caption{Dark and source RHN masses  
that satisfy the resonance condition for two phase transition temperatures. $N_S$ are relativistic in the region between the vertical lines.
}
\label{fig:mass_regime}
\end{figure}
In Fig.~\ref{fig:mass_regime} we show such regions for two values of $T_{\star}$.  
From Eq.~(\ref{dmeq}), if the resonance is crossed quickly enough that the expansion and the variation of 
the off-diagonal term $\Delta H_{\rm mix}(t)$ can be neglected during the crossing, and if the decoherence rate 
$\Gamma_{\rm dec} = (1/2)\,\Gamma_{\rm prod}$ is negligible compared to 
$(d\Delta \widetilde{M}^2(t)/dt)_{\rm res}/(6\,T_{\star})$, then the  $N_{\rm D}$ abundance is very well
described by the Landau-Zener (LZ) formula~\cite{ad,unified},
\be
N^{\rm res}_{N_{\rm D}} \simeq 12\pi \, N_{N_{\rm S}}(T_{\star}) \, T_{\star}  \, 
\left. {\Delta H^2_{\rm mix} \over d \D\widetilde{M}^2/dt} \right|_{t_{\rm res}} \,  .
\ee
An LZ description was also used in the case of a standard cosmological expansion without 
a phase transition in Ref.~\cite{densitym}.  In that case the bare mass $M_{\rm S}$ is constant, 
and for $M_{\rm S} \gg M_{\rm D}$, LZ 
greatly overestimates $N_{\rm D}$ production because at the resonance, $\D M^2 /(2E) \ll H$, so that
RHN oscillations simply do not have time to develop and 
LZ breaks down~\cite{analytical}. In our case, $\D M^2 /(2E) \gg H, v_{\rm w}/\D_{\rm w}$, must be satisfied  to ensure that $N_S$ has enough time to oscillate into $N_D$. 
We set  $v_{\rm w}/\D_{\rm w} = T_{\star}/50$ so that LZ is valid 
and the source RHNs are ultra-relativistic at the
resonance. This holds for a narrow range, $T_{\star}/2 \lesssim M_{\rm S} \lesssim T_{\star}$, between the vertical lines
in Fig.~\ref{fig:mass_regime} for two values of $T_{\star}$.
We solved the density matrix equation numerically to verify that LZ works well within this range, 
so long as the resonance is also crossed quickly enough. One might consider enhancing DM production by 
making $(d \D \widetilde{M}^2/dt)_{\rm res}$ smaller, 
by taking a smaller $v_{\rm w}$, but too small a value of $v_{\rm w}$ invalidates 
LZ which relies on a first order Taylor expansion 
about $t_{\rm res}$. We take $v_{\rm w} =0.9$ and $0.95$ to guarantee
  the validity of LZ.  

\begin{figure}
        \psfig{file=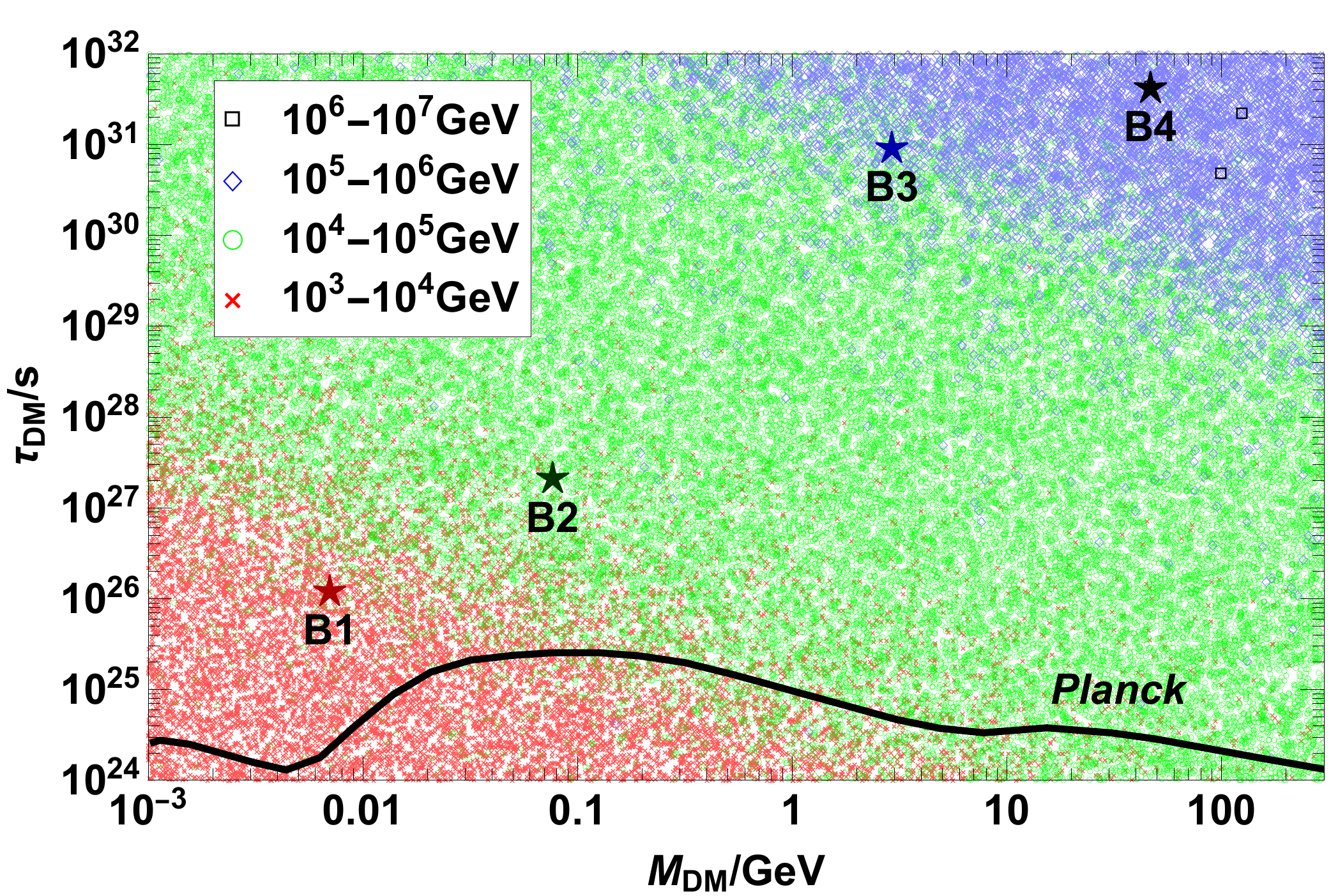,height=50mm,width=80mm}
        \label{fig:gull}
\caption{Points that  reproduce the measured DM abundance.
All points are generated for $\overline{m}_\a = 0.125\,m_{\rm sol} \simeq 1\,{\rm meV}$,  $v_{\rm w}/\Delta_{\rm w} = T_{\star}/50$, and for 
$T_{\star}$ within the five ranges distinguished by the indicated color code. 
The stars mark four of the  five benchmark points in Table~\ref{tab:samples}; B5 has a DM lifetime beyond the range of the figure.  
The black curve indicates the current lower bound on the DM lifetime from {\it Planck} data.
}
\label{fig:scatterplot}
\end{figure}

We calculate $(d \D \widetilde{M}^2/dt)_{\rm res}$, and arrive at the 
dark RHN abundance produced at the resonance:
\be
N_{N_{\rm D}}^{\rm res} \simeq {N_{N_{\rm S}}(T_{\star}) \, (\pi / 48)\,(\D_{\rm w} / v_{\rm w})\, T_{\star}^5 \,
\over \,\widetilde{\Lambda}^2 \, M^2_{\rm S}(r,t_{\rm res}) \, [1-M_{\rm S}(r,t_{\rm res})/M_{\rm S}]} \,   ,
\ee
However, $N_{\rm D}$ cannot be DM, since it  cannot be long-lived enough to simultaneously
satisfy various astrophysics/cosmology constraints (described below) and reproduce the observed DM abundance, $\Omega_{\rm DM}h^2 = 0.11933 \pm 0.00273$~\cite{planck18} at $3\s$. 
Indeed, even at zero temperature $N_D$ decays  because of its mixing with $N_S$. 
We require that $N_{\rm D}$ dominantly decays to $N_{\rm DM}$ via $N_{\rm D} \ra N_{\rm DM} + 2 A$, where $A=W^\pm,Z,\,{\rm{Higgs}}$.
We will find its lifetime, $\tau_{\rm D} \lesssim 10^{-16}\,{\rm s}$.
The $N_{\rm DM}$ abundance is  given by 
\be\label{OmegaNDM}
\O_{N_{\rm DM}}\,h^2 = 1.0875 \times 10^6 \,  N_{N_{\rm D}}^{\rm res}\,{M_{\rm DM} \over {\rm GeV}} \,   .
\ee
However,  $N_{\rm D}$ can mediate the decay of $N_{\rm DM}$, thereby placing a lower bound 
on its lifetime $\tau_{\rm DM} \sim  \widetilde{\L}^{2}$, in contrast to the much shorter $N_{\rm D}$ lifetime which scales as $\widetilde{\L}$.  
For DM masses  below $\sim 300\,{\rm GeV}$, the dominant decay
mode is $N_{\rm D} \ra \nu \, {\ell}^+_{\a}{\ell}^-_{\a} \, (\a =e,\mu,\tau)$ with a rate,  %%%%%%%%%%%%%%%%%%%%%%%%%%%%%%%%%%%%%%%
\be
\G_{N_{\rm DM}\ra \nu \, {\ell}^+_{\a}{\ell}^-_{\a} } = {(\theta^{\rm D-S}_{\L 0} \,  \theta^{\rm DM-D}_{\L 0})^2 \, \over 96\,\pi^3} \, {\overline{m}_\a \over M_{\rm S}}
\, G^2_{\rm F} \, M_{\rm DM}^5 \,  ,
\ee
where $\theta^{\rm D-S (DM-D)}_{\L 0} = 2v^2/(\widetilde{\L}\,(M_{\rm S (D)}-M_{\rm D(DM)}))$ 
is the mixing angle between the dark (dark matter) and source (dark) RHNs at zero temperature~\cite{unified}
and $\overline{m}_\a$ is an effective neutrino mass that depends on $\widetilde{m}_{\rm S}$, an $N_{\rm S}$-$\nu_\alpha$ Yukawa coupling, and SM gauge couplings. 
A robust lower bound, $\tau_{\rm DM} \gtrsim 10^{25}\,{\rm s}$, is
provided by CMB anisotropy data via changes to the ionization and temperature history~\cite{taudmlb}  
(see the {\em Planck} labelled black curve in Fig.~\ref{fig:scatterplot}). 
Data from diffuse $X$-ray and $\gamma$-ray observations 
also place a lower bound $\tau_{\rm DM} \gtrsim 10^{25}\, {\rm s}$~\cite{Essig:2013goa}
(not shown in Fig.~\ref{fig:scatterplot}) in the range 0.1~GeV--10~GeV.  
The points in Fig.~\ref{fig:scatterplot} correspond to sets of parameter values that reproduce the measured DM abundance and respect the
experimental constraints from decays. They are obtained 
by scanning over $M_{\rm DM}$,  $M_{\rm S}$, $T_{\star}$ (uniform in logarithm) and $\tau_{\rm DM}$  (instead of $\widetilde{\L}$).
The solutions are divided into five subsets corresponding to the
five indicated ranges of $T_{\star}$. There is a  tendency for the value of $M_{\rm DM}$
to increase with $T_{\star}$, which results from the resonance condition Eq.~(\ref{MDMMS}) that, as we have seen from Fig.~\ref{fig:mass_regime}, 
imposes a relation between $M_{\rm DM}, M_{\rm S}$ and $T_{\star}$. 

{\bf 4.~Spectrum of gravitational waves}. 
We now estimate the potential for current and future GW interferometers to detect the stochastic GW background produced by the strong first-order cosmological phase transition of $\eta$. We consider the case of non-runaway bubbles~\cite{Caprini:2015zlo,Caprini:2019egz}, 
in which the scalar field contribution to GWs is ignored, and conservatively assume that only 5\% of the bulk motion of the bubble walls is converted into vorticity, as supported by numerical results~\cite{Hindmarsh:2015qta}. 
Then, the contribution from  sound waves is the dominant source of the GW background in most of the frequency range. 
The GW spectrum generated by sound waves during the phase transition is determined by four parameters: 
the phase transition temperature $T_\star$; 
the bubble wall velocity $v_{\rm w}$; 
the ratio of the vacuum energy density released in the transition to the  energy density of the radiation bath $\alpha$; 
the fraction $\beta/H_\star$, where $H_\star$ is the expansion rate at $T_\star$ and $\beta$ is (approximately) the inverse 
of the time duration of the phase transition. The last two parameters are irrelevant for DM genesis. 
We only consider values of $\alpha$ that permit $v_{\rm w}$ much larger than $c_s(1+\sqrt{2\alpha})$, where $c_s=1/\sqrt{3}$ is the sound speed
in the plasma~\cite{Caprini:2019egz}. In this regime the analytical results and fits provided  in Ref.~\cite{Caprini:2015zlo} can be employed.  

\begin{figure}
        \psfig{file=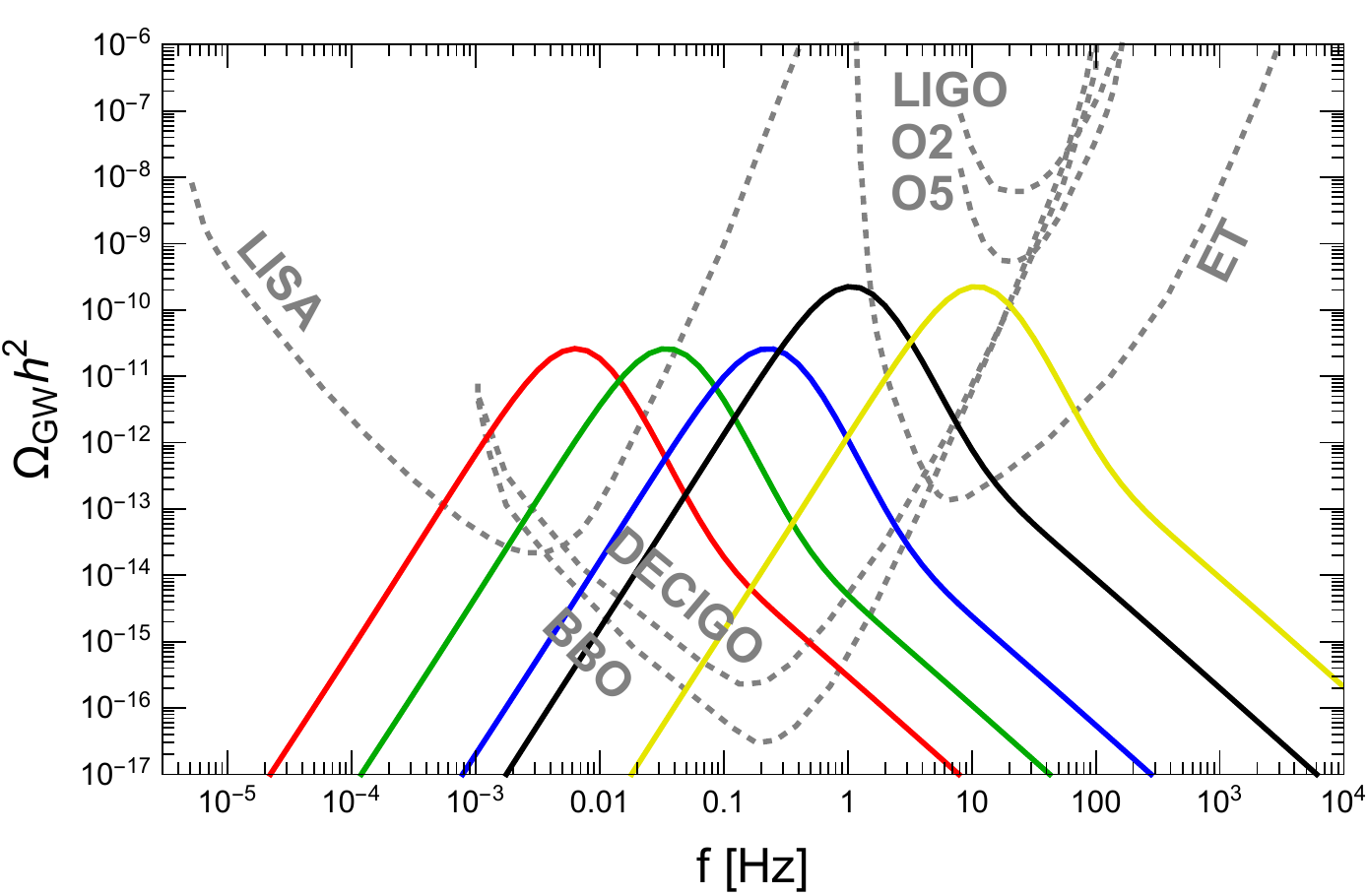,height=50mm,width=80mm}
\caption{Predicted spectrum of GWs for the five benchmark points (from left to right) in Table~\ref{tab:samples}.}
\label{fig:gws}
\end{figure}

In Fig.~\ref{fig:gws} we display the GW spectrum for our five
benchmark points in Table~\ref{tab:samples}. Sensitivities of the LIGO O2 \& O5 observing runs~\cite{Aasi:2013wya}, LISA~\cite{Caprini:2015zlo,Auclair:2019wcv}, 
ET~\cite{Hild:2010id}, BBO~\cite{Yagi:2011wg} and DECIGO~\cite{Kawamura:2019jqt} are shown for comparison.  TianQin~\cite{Luo:2015ght}
and Taiji~\cite{Guo:2018npi} have sensitivities similar to LISA. 
As shown in Fig.~\ref{fig:scatterplot}, successful DM genesis requires $T_\star \gtrsim 10^4$~GeV for the bulk of the points. 
LISA can test this regime. LIGO, ET, BBO and DECIGO might be able to test a high temperature phase transition, $T_\star \sim 10^7\,{\rm GeV}-10^8$~GeV,
if $\alpha$ takes a relatively large value. For a five-year LISA mission with a 75\% duty cycle, we find the signal-to-noise ratio for B1-B5 to be ${\cal{O}}(10^5,10^4,10^2,10,10^{-2})$, respectively. Since in all cases $M_{\rm S}$ is larger than 300~GeV, leptogenesis is viable. 
\begin{table}[t]
\begin{center}
\begin{tabular}{ lcccccccc } 
 \hline
    & ${T_\star\over {\rm PeV}}$ & ${\tau_{\rm DM}\over 10^{26}{\rm s}}$ & ${M_{\rm S}\over {\rm TeV}}$ & ${M_{\rm D}\over {\rm TeV}}$ & ${M_{\rm DM}\over {\rm GeV}}$ & $v_{\rm w}$ & $\alpha$ & $\frac{\beta}{H_\star}$  \\\hline
B1 & $3\!\cdot\! 10^{-3}$ & 1.219 & $1.57$ & $0.567$ & $7\!\cdot\! 10^{-3}$ & 0.90 & 0.10 & 10 \\
B2 & $0.016$ & 21.26 & $12.9$ & $7.72$ & 0.077 & 0.90 & 0.10 & 10 \\
B3 & $0.106$ & $9.25\! \cdot \! 10^4$ & $93.3$ & $72.6$ & 2.92 & 0.90 & 0.10 & 10 \\
B4 & $1.052$ & $4.24 \! \cdot \! 10^5$ & $666$ & $666$ & 46.69 & 0.95 & 0.15 & 5 \\
B5 & $10.75$ & $4.69\! \cdot \! 10^{17}$ & $8.7\! \cdot\! 10^3$ & $5.3\! \cdot\! 10^3$ & 175.8 & 0.95 & 0.15 & 5 \\
 \hline
\end{tabular}
\end{center}
\caption{Benchmark points obtained for $v_{\rm w}/\D_{\rm w} = T_{\star}/50$. 
}
\label{tab:samples} 
\end{table}

{\bf 4.~Final remarks.} We introduced a novel model in which DM and neutrino masses are generated during  
a phase transition of a scalar field. The phase transition needs to be strongly first order for efficient DM genesis, which has the welcome
feature of the associated production of GWs with an intensity measurable 
at the planned GW interferometers, LISA, BBO, DECIGO, ET, TianQin and Taiji.
We  find solutions for DM masses within a range $1\,{\rm MeV}$--$300\,{\rm GeV}$ with a clear tendency of the
temperature of the phase transition to increase with the dark matter mass, though the dependence also involves 
other parameters such as the mass of the source RHN. Interestingly,  our solutions have source RHNs heavier than $~300\,{\rm GeV}$
which allows successful resonant leptogenesis~\cite{resonantlep} with two RHNs \cite{densitym}.
We have also seen that the decays of the dark RHNs offer an additional way to test the scenario through observations of CMB anisotropies and/or the diffuse X- and $\gamma$-ray backgrounds. 
We have not thoroughly studied the dependence of our results on all parameters and have 
fixed some of them to reasonable values, e.g., $g_{\eta} =1/6$, $v_{\rm w} =T_{\star}$ and $\widetilde{m}_{\rm S}/m_{\rm sol}=1$.
However, we verified that our results are insensitive to small variations of the first two parameters.
We restricted ourselves to values of $M_{\rm S} \simeq T_\star$ so as to use the LZ approximation. 
Numerical density matrix calculations should determine the full range of allowed values for $M_{\rm S}/T_{\star}$.  
We also assumed a specific bubble wall profile with a fixed value of the  
bubble width and velocity.  A thorough exploration of the allowed values of all parameters, including the possibility of DM masses above 300~GeV, will be presented in a forthcoming paper 
together with some variants of the model~\cite{forthcoming}. 
However, our results clearly prove the viability of the proposed mechanism and how DM genesis,
 although escaping traditional direct and collider searches, will be tested in the future  with 
 cosmology, cosmic rays and GWs.

{\bf{Acknowledgments.} }We thank L.~Bian, Z.~Liu, K.~Farrag and R.~Samanta, for useful discussions. 
DM thanks the University of Southampton and the Aspen Center for
Physics (which is supported by U.S. NSF Grant No. PHY-1607611) for their hospitality while this work was in progress. PDB thanks INT at the University of Washington for its hospitality and the U.S. DOE for partial support during the completion of this work. 
PDB and YLZ acknowledge financial support from the STFC Consolidated Grant L000296/1. 
This project has received funding/support from the European Union Horizon 2020 research and innovation 
programme under the Marie Sk\l{}odowska-Curie grant agreements number 690575 and  674896. DM is supported in
part by U.S. DOE Grant No. de-sc0010504.


\begin{thebibliography}{99}


\bibitem{bertonehooper}
 G.~Bertone and D.~Hooper,
% {\em History of dark matter},
  Rev.\ Mod.\ Phys.\  {\bf 90} (2018) no.4,  045002
%  doi:10.1103/RevModPhys.90.045002
  [arXiv:1605.04909 [astro-ph.CO]];
  %%CITATION = doi:10.1103/RevModPhys.90.045002;%%
  %210 citations counted in INSPIRE as of 09 Dec 2019
%for a general up-to-date discussion on dark matter see
%for example Chapter 17 in 
P.~Di Bari, {\em Cosmology and the early universe}, CRC Press, May 2018
(ISBN 9781498761703).


\bibitem{ad}
A.~Anisimov and P.~Di Bari,
%  {\em Cold Dark Matter from heavy Right-Handed neutrino mixing},
  Phys.\ Rev.\ D {\bf 80} (2009) 073017
%  doi:10.1103/PhysRevD.80.073017
  [arXiv:0812.5085 [hep-ph]].
  %%CITATION = doi:10.1103/PhysRevD.80.073017;%%
  %25 citations counted in INSPIRE as of 07 May 2019

\bibitem{anisimov}
  A.~Anisimov, %{\em Majorana Dark Matter},
  % doi:10.1142/9789812770288\_0058, 
  hep-ph/0612024.
  %%CITATION = doi:10.1142/9789812770288_0058;%%
  %6 citations counted in INSPIRE as of 05 Sep 2019



\bibitem{unified}
P.~Di Bari, P.~O.~Ludl and S.~Palomares-Ruiz,
% {\em Unifying leptogenesis, dark matter and high-energy neutrinos with right-handed neutrino mixing via Higgs portal},
  JCAP {\bf 1611} (2016) no.11,  044
%  doi:10.1088/1475-7516/2016/11/044
  [arXiv:1606.06238 [hep-ph]].
  %%CITATION = doi:10.1088/1475-7516/2016/11/044;%%
  %33 citations counted in INSPIRE as of 07 May 2019

\bibitem{densitym}
  P.~Di Bari, K.~Farrag, R.~Samanta and Y.~L.~Zhou,
%  {\em Density matrix calculation of the dark matter abundance in the Higgs induced right-handed neutrino mixing model},
  arXiv:1908.00521 [hep-ph].
  %%CITATION = ARXIV:1908.00521;%%

\bibitem{yelingjessicasilvia}
 S.~Pascoli, J.~Turner and Y.~L.~Zhou,
  %{\em Baryogenesis via leptonic CP-violating phase transition},
  Phys.\ Lett.\ B {\bf 780} (2018) 313
%  doi:10.1016/j.physletb.2018.03.011
  [arXiv:1609.07969 [hep-ph]].
  %%CITATION = doi:10.1016/j.physletb.2018.03.011;%%
  %8 citations counted in INSPIRE as of 07 May 2019


\bibitem{witten}
E.~Witten,
% {\em Cosmic Separation of Phases},
  Phys.\ Rev.\ D {\bf 30} (1984) 272.
%  doi:10.1103/PhysRevD.30.272
  %%CITATION = doi:10.1103/PhysRevD.30.272;%%
  %2365 citations counted in INSPIRE as of 15 Jan 2020

\bibitem{hogan}
C.~Hogan,
%  {\em Gravitational radiation from cosmological phase transitions},
  Mon.\ Not.\ Roy.\ Astron.\ Soc.\  {\bf 218} (1986) 629.
  %%CITATION = MNRAA,218,629;%%
  %138 citations counted in INSPIRE as of 17 Jan 2020


\bibitem{tw}
M.~Turner and F.~Wilczek,
%  {\em Inflationary axion cosmology},
  Phys.\ Rev.\ Lett.\  {\bf 66} (1991) 5.
% doi:10.1103/PhysRevLett.66.5
  %%CITATION = doi:10.1103/PhysRevLett.66.5;%%
  %170 citations counted in INSPIRE as of 17 Jan 2020


\bibitem{kkt}
M.~Kamionkowski, A.~Kosowsky and M.~S.~Turner,
%  {\em Gravitational radiation from first order phase transitions},
  Phys.\ Rev.\ D {\bf 49} (1994) 2837
%  doi:10.1103/PhysRevD.49.2837
  [astro-ph/9310044].
  %%CITATION = doi:10.1103/PhysRevD.49.2837;%%
  %396 citations counted in INSPIRE as of 17 Jan 2020

\bibitem{Cohen:2008nb}
  T.~Cohen, D.~E.~Morrissey and A.~Pierce,
%  {\em Changes in Dark Matter Properties After Freeze-Out},
  Phys.\ Rev.\ D {\bf 78} (2008) 111701
%  doi:10.1103/PhysRevD.78.111701
  [arXiv:0808.3994 [hep-ph]].
  %%CITATION = doi:10.1103/PhysRevD.78.111701;%%
  %33 citations counted in INSPIRE as of 17 Jan 2020


\bibitem{Falkowski:2012fb}
A.~Falkowski and J.~M.~No,
%  {\em Non-thermal Dark Matter Production from the Electroweak Phase Transition: Multi-TeV WIMPs and 'Baby-Zillas'},
  JHEP {\bf 1302} (2013) 034
%  doi:10.1007/JHEP02(2013)034
  [arXiv:1211.5615 [hep-ph]].
  %%CITATION = doi:10.1007/JHEP02(2013)034;%%
  %11 citations counted in INSPIRE as of 15 Jan 2020

\bibitem{Huang:2017kzu}
  F.~P.~Huang and C.~S.~Li,
%  {\em Probing the baryogenesis and dark matter relaxed in phase transition by gravitational waves and colliders},
  Phys.\ Rev.\ D {\bf 96} (2017) no.9,  095028
%  doi:10.1103/PhysRevD.96.095028
  [arXiv:1709.09691 [hep-ph]].
  %%CITATION = doi:10.1103/PhysRevD.96.095028;%%
  %14 citations counted in INSPIRE as of 16 Jan 2020

\bibitem{Bian:2018mkl}
  L.~Bian and Y.~L.~Tang,
%  {\em Thermally modified sterile neutrino portal dark matter and gravitational waves from phase transition: The Freeze-in case},
  JHEP {\bf 1812} (2018) 006
%  doi:10.1007/JHEP12(2018)006
  [arXiv:1810.03172 [hep-ph]].
  %%CITATION = doi:10.1007/JHEP12(2018)006;%%
  %12 citations counted in INSPIRE as of 15 Jan 2020

\bibitem{Bai:2018dxf}
  Y.~Bai, A.~J.~Long and S.~Lu,
%  {\em Dark Quark Nuggets},
  Phys.\ Rev.\ D {\bf 99} (2019) no.5,  055047
 % doi:10.1103/PhysRevD.99.055047
  [arXiv:1810.04360 [hep-ph]].
  %%CITATION = doi:10.1103/PhysRevD.99.055047;%%
  %19 citations counted in INSPIRE as of 16 Jan 2020

\bibitem{murayama}
 E.~Hall, T.~Konstandin, R.~McGehee and H.~Murayama,
%  {\em Asymmetric Matters from a Dark First-Order Phase Transition},
  arXiv:1911.12342 [hep-ph].
  %%CITATION = ARXIV:1911.12342;%%
  %1 citations counted in INSPIRE as of 09 Dec 2019

\bibitem{kopp}
M.~Baker, J.~Kopp, A.~Long,
%  {\em Filtered Dark Matter at a First Order Phase Transition},
  arXiv:1912.02830 [hep-ph].
  %%CITATION = ARXIV:1912.02830;%%
  
 

\bibitem{Chway:2019kft}
  D.~Chway, T.~H.~Jung and C.~S.~Shin,
%  {\em Dark matter filtering-out effect during a first-order phase transition},
  arXiv:1912.04238 [hep-ph].
  %%CITATION = ARXIV:1912.04238;%%
  %1 citations counted in INSPIRE as of 24 Jan 2020
  
   \bibitem{tseng}
  D.~Marfatia, P.~Y.~Tseng,
%``Gravitational wave signals of dark matter freeze-out,''
[arXiv:2006.07313 [hep-ph]].

\bibitem{fy}
M.~Fukugita, T.~Yanagida,
%  {\em Baryogenesis Without Grand Unification},
  Phys.\ Lett.\ B {\bf 174} (1986) 45.
%  doi:10.1016/0370-2693(86)91126-3
  %%CITATION = doi:10.1016/0370-2693(86)91126-3;%%
  %3164 citations counted in INSPIRE as of 27 May 2019

\bibitem{john}
  P.~John,
%  {\em Bubble wall profiles with more than one scalar field: A Numerical approach},
  Phys.\ Lett.\ B {\bf 452} (1999) 221
%  doi:10.1016/S0370-2693(99)00272-5
  [hep-ph/9810499].
  %%CITATION = doi:10.1016/S0370-2693(99)00272-5;%%
  %48 citations counted in INSPIRE as of 09 Dec 2019

\bibitem{analytical}
P.~Di Bari, R.~Samanta, Y.L.~Zhou, in preparation. 


\bibitem{planck18}
 Y.~Akrami {\it et al.} [Planck Collaboration],
% {\em Planck 2018 results. I. Overview and the cosmological legacy of Planck},
  arXiv:1807.06205 [astro-ph.CO].
  %%CITATION = ARXIV:1807.06205;%%
  %262 citations counted in INSPIRE as of 10 Jan 2020


\bibitem{taudmlb}
T.~R.~Slatyer and C.~L.~Wu,
%  {\em General Constraints on Dark Matter Decay from the Cosmic Microwave Background},
  Phys.\ Rev.\ D {\bf 95} (2017) no.2,  023010
%  doi:10.1103/PhysRevD.95.023010
  [arXiv:1610.06933 [astro-ph.CO]].
  %%CITATION = doi:10.1103/PhysRevD.95.023010;%%
  %95 citations counted in INSPIRE as of 10 Jan 2020

\bibitem{Essig:2013goa}
  \mbox{R.~Essig, E.~Kuflik, S.~McDermott, T.~Volansky, K.~Zurek},
% {\em Constraining Light Dark Matter with Diffuse X-Ray and Gamma-Ray Observations},
  JHEP {\bf 1311} (2013) 193
 % doi:10.1007/JHEP11(2013)193
  [arXiv:1309.4091 [hep-ph]].
  %%CITATION = doi:10.1007/JHEP11(2013)193;%%
  %148 citations counted in INSPIRE as of 18 Jan 2020

\bibitem{Caprini:2015zlo} 
  C.~Caprini {\it et al.},
%  {\em Science with the space-based interferometer eLISA. II: Gravitational waves from cosmological phase transitions},
  JCAP {\bf 1604}, 001 (2016)
 % doi:10.1088/1475-7516/2016/04/001
  [arXiv:1512.06239 [astro-ph.CO]].
  %%CITATION = doi:10.1088/1475-7516/2016/04/001;%%
  %290 citations counted in INSPIRE as of 10 Jan 2020
  
  \bibitem{Caprini:2019egz} 
  C.~Caprini {\it et al.},
% {\em Detecting gravitational waves from cosmological phase transitions with LISA: an update},
  arXiv:1910.13125 [astro-ph.CO].
  %%CITATION = ARXIV:1910.13125;%%

%\cite{Hindmarsh:2015qta}
\bibitem{Hindmarsh:2015qta} 
  M.~Hindmarsh, S.~J.~Huber, K.~Rummukainen and D.~J.~Weir,
%  {\em Numerical simulations of acoustically generated gravitational waves at a first order phase transition},
  Phys.\ Rev.\ D {\bf 92}, no. 12, 123009 (2015)
%  doi:10.1103/PhysRevD.92.123009
  [arXiv:1504.03291 [astro-ph.CO]].
  %%CITATION = doi:10.1103/PhysRevD.92.123009;%%
  %139 citations counted in INSPIRE as of 10 Jan 2020

%\cite{Aasi:2013wya}
\bibitem{Aasi:2013wya} 
  B.~P.~Abbott {\it et al.} [KAGRA and LIGO Scientific and VIRGO Collaborations],
%  {\em Prospects for Observing and Localizing Gravitational-Wave Transients with Advanced LIGO, Advanced Virgo and KAGRA},
  Living Rev.\ Rel.\  {\bf 21}, no. 1, 3 (2018)
 %  doi:10.1007/s41114-018-0012-9, 10.1007/lrr-2016-1
  [arXiv:1304.0670 [gr-qc]].
  %%CITATION = doi:10.1007/s41114-018-0012-9, 10.1007/lrr-2016-1;%%
  %831 citations counted in INSPIRE as of 10 Jan 2020


%\cite{Auclair:2019wcv}
\bibitem{Auclair:2019wcv} 
  P.~Auclair {\it et al.},
%  {\em Probing the gravitational wave background from cosmic strings with LISA},
  arXiv:1909.00819 [astro-ph.CO].
  %%CITATION = ARXIV:1909.00819;%%
  %9 citations counted in INSPIRE as of 10 Jan 2020


%\cite{Hild:2010id}
\bibitem{Hild:2010id} 
  S.~Hild {\it et al.},
%  {\em Sensitivity Studies for Third-Generation Gravitational Wave Observatories},
  Class.\ Quant.\ Grav.\  {\bf 28}, 094013 (2011)
%  doi:10.1088/0264-9381/28/9/094013
  [arXiv:1012.0908 [gr-qc]].
  %%CITATION = doi:10.1088/0264-9381/28/9/094013;%%
  %222 citations counted in INSPIRE as of 10 Jan 2020


%\cite{Yagi:2011wg}
\bibitem{Yagi:2011wg} 
  K.~Yagi and N.~Seto,
%  {\em Detector configuration of DECIGO/BBO and identification of cosmological neutron-star binaries},
  Phys.\ Rev.\ D {\bf 83}, 044011 (2011)
  Erratum: [Phys.\ Rev.\ D {\bf 95}, no. 10, 109901 (2017)]
 % doi:10.1103/PhysRevD.95.109901, 10.1103/PhysRevD.83.044011
  [arXiv:1101.3940 [astro-ph.CO]].
  %%CITATION = doi:10.1103/PhysRevD.95.109901, 10.1103/PhysRevD.83.044011;%%
  %99 citations counted in INSPIRE as of 10 Jan 2020
 
 \bibitem{Kawamura:2019jqt}
  S.~Kawamura {\it et al.} [DECIGO Collaboration],
%  {\em Primordial gravitational wave and DECIGO},
  PoS KMI {\bf 2019} (2019) 019.
%  doi:10.22323/1.356.0019
  %%CITATION = doi:10.22323/1.356.0019;%%
 
  
  
%\cite{Luo:2015ght}
\bibitem{Luo:2015ght} 
  J.~Luo {\it et al.} [TianQin Collaboration],
%  {\em TianQin: a space-borne gravitational wave detector},
  Class.\ Quant.\ Grav.\  {\bf 33}, no. 3, 035010 (2016)
%  doi:10.1088/0264-9381/33/3/035010
  [arXiv:1512.02076 [astro-ph.IM]].
  %%CITATION = doi:10.1088/0264-9381/33/3/035010;%%
  %147 citations counted in INSPIRE as of 10 Jan 2020

%\cite{Guo:2018npi}
\bibitem{Guo:2018npi} 
  W.~H.~Ruan, Z.~K.~Guo, R.~G.~Cai and Y.~Z.~Zhang,
%  {\em Taiji Program: Gravitational-Wave Sources},
  arXiv:1807.09495 [gr-qc].
  %%CITATION = ARXIV:1807.09495;%%
  %31 citations counted in INSPIRE as of 10 Jan 2020

\bibitem{resonantlep}
A.~Pilaftsis and T.~E.~J.~Underwood,
%  {\em Resonant leptogenesis},
  Nucl.\ Phys.\ B {\bf 692} (2004) 303
% doi:10.1016/j.nuclphysb.2004.05.029
  [hep-ph/0309342].
  %%CITATION = doi:10.1016/j.nuclphysb.2004.05.029;%%
  %672 citations counted in INSPIRE as of 16 Jan 2020



\bibitem{forthcoming}
P.~Di Bari, D.~Marfatia, Y.L.~Zhou, in preparation. 


\end{thebibliography}
\end{document}